# Contiguity is limited in free recall

Eugen Tarnow

18-11 Radburn Road

Fair Lawn, NJ 07410

etarnow@avabiz.com

## Abstract:

Does the "law of contiguity" apply to free recall? I find that conditional response probabilities, often used as evidence for contiguity in free recall, have been displayed insufficiently, limiting the distance from the last item recalled, and only averaged over all items. When fully expanding the x-axis and examining the conditional response probabilities separately for the beginning, middle and end of the presented item list, I show that the law of contiguity applies only locally, even then only sometimes, and breaks down globally.

**Introduction**

Free recall stands out as one of the great unsolved mysteries of modern psychology. Words in a list are displayed or read to subjects who are then asked to retrieve the words. It is one of the simplest ways to probe short term memory but the results (Murdock, 1960; Murdock, 1962; Murdock, 1975) have defied explanation. Why do we remember primarily items in the beginning and in the end of the list, but not items in the middle, creating the famous u-shaped curve of probability of recall versus serial position? Why can we remember 50-100 items in cued recall or recognition but only 6-8 items in free recall?

Efforts to understand free recall include efforts to impose some kind of order on the results. In particular, a "law of contiguity" has been suggested to cover the order of item recalls in free recalls. This law "implies that items studied in neighboring list positions serve as more effective retrieval cues for one another than do items studied in remote list positions" (Davis et al, 2008). The evidence presented (Kahana, 1996; Howard and Kahana, 2002; Howard, Kahana and Zaromb, 2002; Sederberg, Howard and Kahana, 2008; Davis et al, 2008; Miller, Weidemann and Kahana, 2012) includes conditional response probability (CRP) curves such as the one in Fig. 1. (The data in this contribution come from the 40-1 list of Murdock (1962) in which 40 words are read to the subjects at a rate 1 item per second. I have chosen to display the conditional probability as a function of the distance from the last item instead of lag: -1 means the previous item and +1 is the subsequent item. The distance from the last item is the negative of the lag.)

The argument for a "law of contiguity" typically reads as follows "the probability of recalling a word from serial position i + lag immediately following a word from serial position i is a decreasing function of | lag |. This contiguity effect exhibits a forward bias, with associations being stronger in the forward direction than in the backward direction" (David et al, 2008). Even stronger: "To our knowledge there is no published failure to find a temporal contiguity effect in a standard free recall task" (Healey & Kahana, 2014). However, limiting the plot to 4-5 previous and subsequent items and only displaying the average over all items, prevents the reader from seeing the global picture. The global picture is quite different.

In Fig. 2a is shown the CRP curve for the first item presented. Clearly, this curve cannot have a forward bias since there is no previous item to cue it. While the second item is more likely to lead to the first item than the third item, thus supporting the "law of contiguity" locally (for small distances between items), the most likely item to lead to the first item is the last item! Indeed, there seems to be an increase in probability of recalling the first item the larger the distance is from the previous item. Globally there are two maximums of the CRP curves, not one: the subsequent item and the last item. If we insist that the "law of contiguity" applies, we would then have to conclude that the first item is somehow related to the last item, leading to a contradiction.

In Fig. 2b is shown the CRP curve for the second item presented. There is a 40% probability that it is cued from the previous item (indicating contiguity), but the subsequent item is less than 5% likely to cue the second item (not indicating contiguity). Globally there are two maxima, the previous item and the last item. The forward bias indicated by the previous item is obliterated by the backward bias indicated by the probability increase leading to the last item.

In Fig. 3a is shown the CRP averaged over nine of the middle items. The item presented just before is the most likely to cue a middle item but this forward bias disappears for subsequent items with roughly equal probability of cueing the middle items and a second local maximum. While there is a local decrease in probability as a function of distance from the last recalled item, this is not true globally. Globally there are two maxima of the CRP curves, the previous and last items, and further distant previous and subsequent items have an equal probability of cueing the middle items. In Fig. 3b is plotted the integrated conditional response probability curve. If only the previous item was important we would have a step function. If only items near the previous and subsequent item were important we would have a rounded step function. But the function is instead a roughly linear function with a small rounded step in the middle corresponding to only about 25-30% of the total.

The occurrence of multiple maxima of the CRP curves occurs also for the last items, see Fig. 4. One global maximum is the previous item, another is the very last item.

The average distance from the last item recalled as function of the presentation position is shown in Fig. 5. It is a monotonically decreasing function with slow variation for the middle item positions. Note the large values of the average distance to the previously recalled item for all but the last items which break the "law of contiguity".

Finally, in Fig. 6 is displayed the CRP averaged over all items for the full x-axis. The average probability decreases for previous items but stays constant or slightly increases for subsequent items. For this average, the law of contiguity and forward bias thus apply locally but not globally.

Why does free recall break the "law of contiguity"? Presumably the most important reason is the presence of boundary conditions which make distinctions between the items: there is a beginning and an end to any presented list.

**Method**

This article makes use of the Murdock (1962) data set (downloaded from the Computational Memory Lab at the University of Pennsylvania (http://memory.psych.upenn.edu/DataArchive). In Table 1 is summarized the experimental processes which generated the data sets used in this paper.

Running head: Contiguity is limited in free recall	6## References

TABLES

| *Work* | *Item types* | *List length and presentation interval* | *Recall interval* | *Subjects* | *Item presentation mode* |
|---|---|---|---|---|---|
| *Murdock (1962)* | *Selection from 4000 most common English words, referred to as the Toronto Word Pool.* | *40 words in a list, each word presented once a second* | *1.5 minutes* | *103 undergraduates* | *Verbal* |

*Table 1. Experimental method that generated the data used in this contribution.*

FIGURES

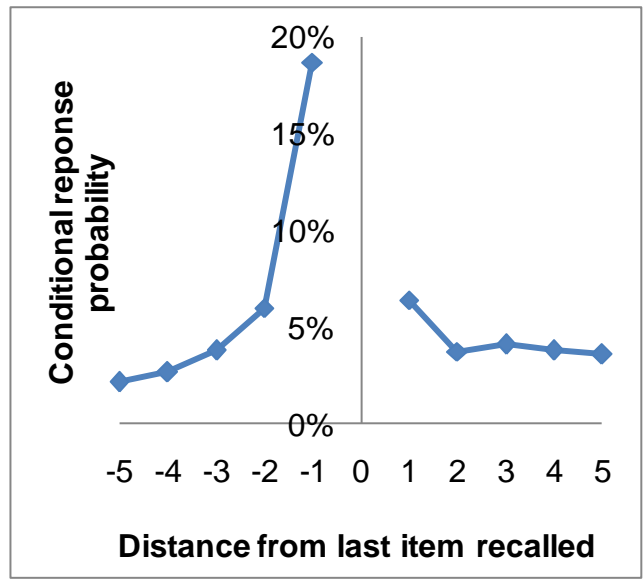

Fig. 1.  Conditional response probability as is commonly plotted.

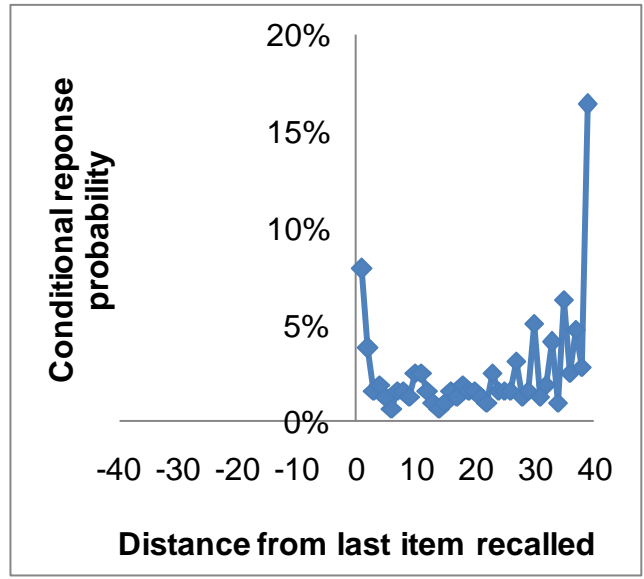

Fig. 2a. Conditional response probability for the first item.

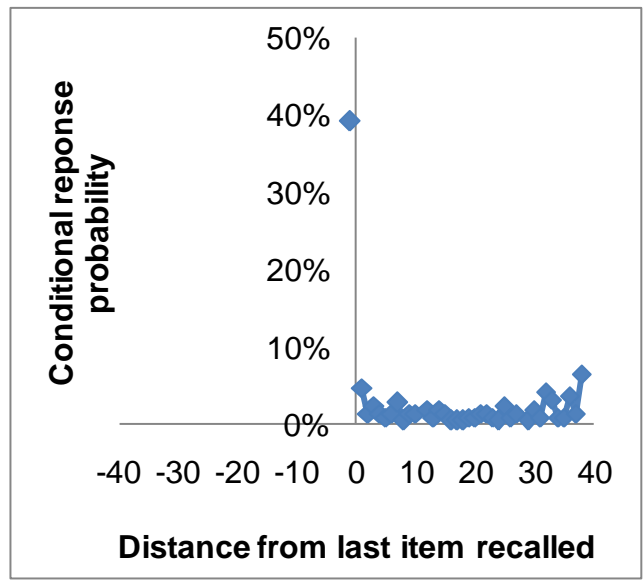

Fig. 2b. Conditional response probability for the second item.

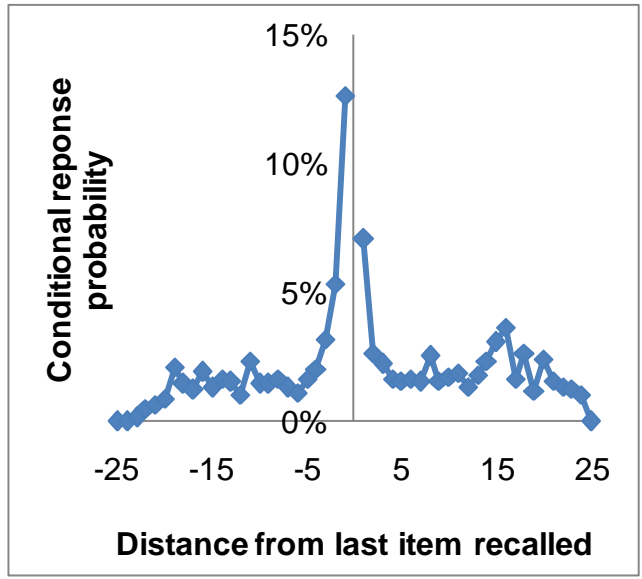

Fig. 3a. Conditional response probability for items presented in the middle of the list.

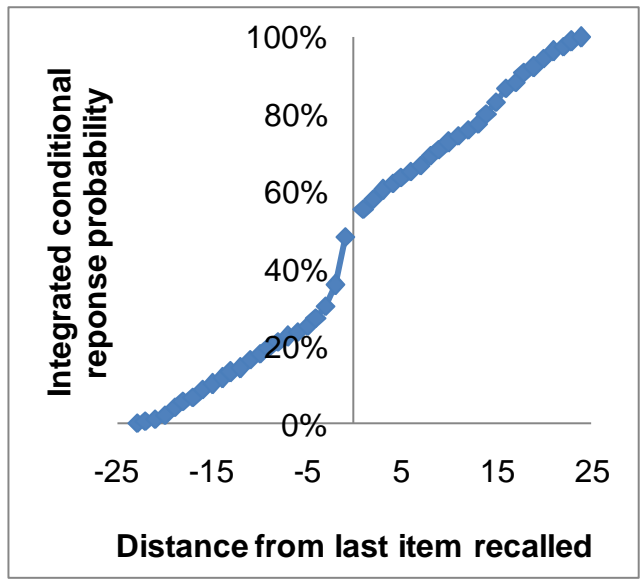

Fig. 3b. Integrated conditional response probability for items presented in the middle of the list.

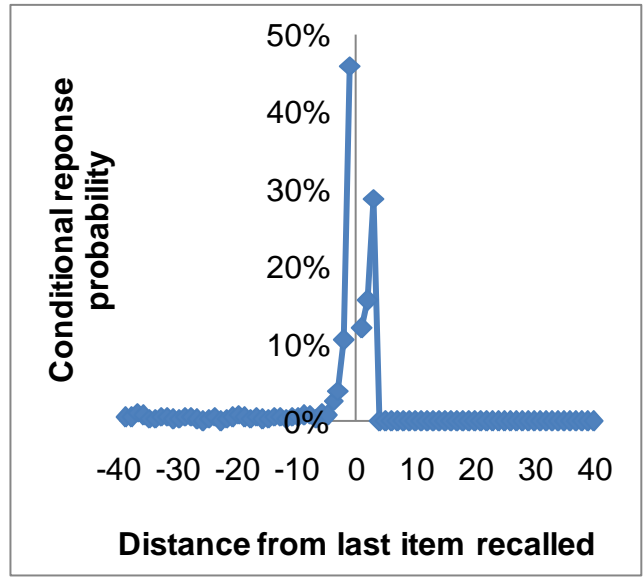

Fig. 4. Conditional response probability averaged over the last four items.

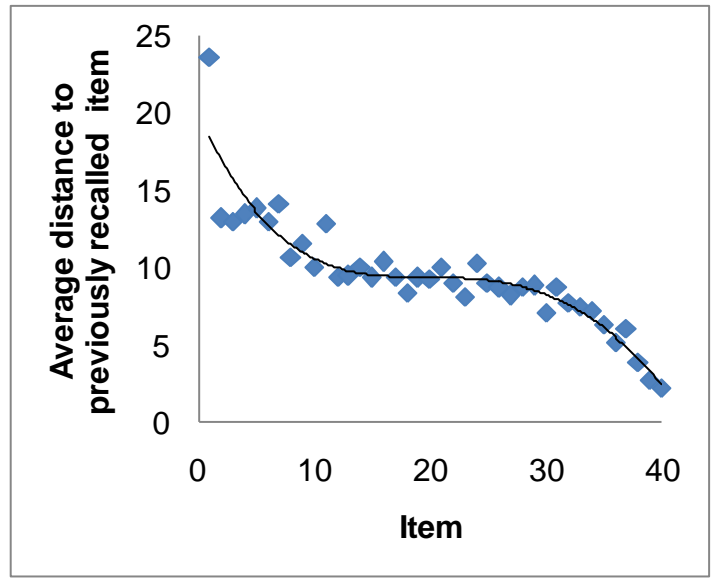

Fig. 5. The average distance between two recalls in Murdock 40-1.

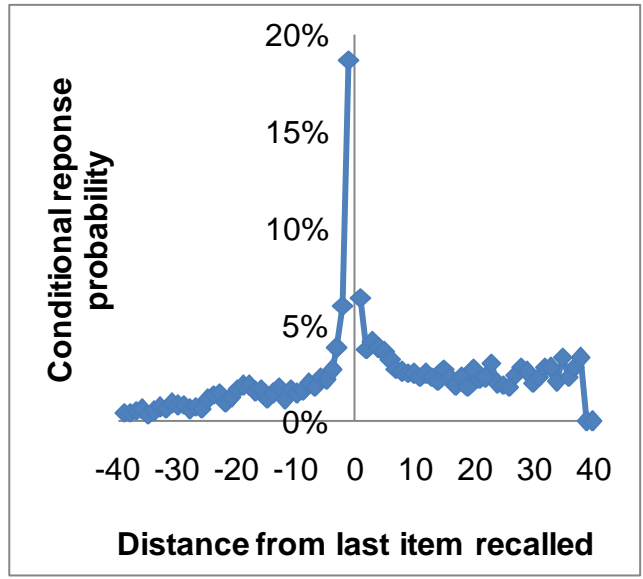

Fig. 6. Complete results for the conditional response probability averaged over all items.